\documentclass[12pt]{article}

\usepackage{amssymb}
\usepackage{amsmath}
\usepackage{graphicx}

\catcode`@=11
\renewcommand{\section}{\@startsection{section}{1}{0pt}{\medskipamount}
{\medskipamount}{\large\bf}}
\numberwithin{equation}{section}
\catcode`@=12

\def\beq{\begin{eqnarray}}    
\def\eeq{\end{eqnarray}}      

\def\ln{\,\mbox{ln}\,}                  
\def\sTr{\,\mbox{sTr}\,}
\def\sDet{\,\mbox{sDet}\,}

\def\im{\textrm{i}}

\def\sfrac#1#2{{\textstyle\frac{#1}{#2}}}
\def\={\ =\ }
\def\und{\qquad\textrm{and}\qquad}

\begin{document}

\begin{titlepage}
\setcounter{page}{0}
\begin{flushright}
ITP--UH--09/13
\end{flushright}

\vskip 1.0cm

\begin{center}

{\LARGE\bf Field-dependent BRST transformations \\[12pt]
           in Yang-Mills theory}

\vspace{18mm}

{\Large Peter M. Lavrov$\,{}^{\dag}$
$^{}\footnote{E-mail: lavrov@tspu.edu.ru}$
\quad and \quad
Olaf Lechtenfeld$\,{}^{\ddag}$
$^{}\footnote{E-mail: lechtenf@itp.uni-hannover.de}$
}

\vspace{8mm}

\noindent ${}^{\dag}${\em
Tomsk State Pedagogical University,\\
Kievskaya St.\ 60, 634061 Tomsk, Russia}

\vspace{4mm}

\noindent ${}^{\ddag}${\em
Institut f\"ur Theoretische Physik and Riemann Center for Geometry and Physics, \\
Leibniz Universit\"at Hannover,
Appelstrasse 2, 30167 Hannover, Germany}

\vspace{20mm}

\begin{abstract}
\noindent
We find an explicit form for the Jacobian of arbitrary field-dependent
BRST transformations in Yang-Mills theory. For the functional-integral
representation of the (gauge-fixed) Yang-Mills vacuum functional,
such transformations merely amount to a precise change in the gauge-fixing
functional. This proves the independence of the vacuum functional under
any field-dependent BRST transformation. We also give a formula for the
transformation parameter functional which generates a prescribed change
of gauge and evaluate it for connecting two arbitrary $R_\xi$ gauges.
\end{abstract}

\end{center}

\vfill


\end{titlepage}


\section{Introduction and summary}

\noindent
In quantum field theory, changing variables in the functional-integral
representation of generating functionals for Greens functions is a major
calculational tool.
In particular, the derivation of Slavnov-Taylor identities
\cite{S,Taylor} in Yang-Mills theory and of Ward identities
in general gauge theories~\cite{BV} utilize field substitutions
related to BRST symmetry~\cite{brs,t}.
In the quantization of dynamical systems with constraints,
the gauge independence of S-matrix elements in the Batalin-Fradkin-Vilkovisky
formalism is proven via a canonical change of variables
pertaining to the Hamiltonian version of BRST symmetry,
with constant as well as with field-dependent parameter~\cite{FV,BV5}.
The Batalin-Vilkovisky formalism for the covariant quantization of
general gauge theories~\cite{BV} also employs a change of functional variables
in a form of field-dependent BRST transformations.

In present paper, we investigate field-dependent BRST transformations
in Yang-Mills theory. They are obtained from standard BRST transformations
by replacing the Grassmann-odd constant parameter~$\lambda$ with
a Grassmann-odd functional~$\Lambda(\phi)$ of the fields~$\phi$ in the theory.
For a change of variables of such type in a functional integral,
we derive a simple explicit form of its Jacobian in terms of the Slavnov variation
of~$\Lambda(\phi)$. By `inverting' this variation, we absorb any such
field-dependent BRST transformation inside the Yang-Mills vacuum functional
into a modification of its gauge-fixing functional.
This proves the independence of the vacuum functional under arbitrary
field-dependent BRST transformations.
Let us turn the question around: Given two different gauges, can one construct
a field-dependent BRST transformation which brings us from one to the other?
Our result gives an explicit formula for the answer.
We demonstrate its use by computing the parameter functional~$\Lambda(\phi)$
which connects any two $R_{\xi}$ gauges (including the Landau gauge).

The paper is organized as follows. In Section~2, we briefly review the salient
features of the Faddeev-Popov quantization of Yang-Mills fields coupled to
arbitrary matter. Section~3 introduces field-dependent BRST transformations
and computes their Jacobian (the supertrace of the Jacobian supermatrix).
In Section 4, we write (the log of) the Jacobian as a Slavnov variation of
a shift in the gauge-fixing functional and solve for $\Lambda(\phi)$ in terms
of the latter, before giving the explicit solution in general and for the case 
of two $R_\xi$ gauges.

We employ the condensed notation of DeWitt~\cite{DeWitt}.
Derivatives with respect to fields are taken from the right.
Left functional derivatives are labeled by a subscript~$l$.

\vspace{0.5cm}

\section{Yang-Mills theory in Faddeev-Popov quantization}

\noindent
Our framework in this paper is Yang-Mills theory of gauge potentials
$A^{a\mu}(x)$ (with Lorentz index $\mu$ and color index $a$),
coupled to some matter fields, such as scalars $\varphi^r(x)$ or
spinors $\psi^s(x)$ (where $r$ and $s$ are gauge group representation
indices). In this section, we introduce our notation and remind the reader
of the basics of the Faddeev-Popov quantization method~\cite{FP}.

For convenience of notation,
let us introduce joint (discrete and continuous) indices
\begin{equation}
i=(x,\mu,a,r,s,\ldots) \und \alpha=(x,a)
\end{equation}
and group the above (physical) fields as
\begin{equation}
\bigl\{A^i\bigr\} = \bigl\{A^{a\mu}(x), \varphi^r(x), \psi^s(x), \ldots\bigr\}
\end{equation}
with Grassmann parities $\varepsilon(A^i)\equiv\varepsilon_i$.
We often abbreviate functional derivatives as $\delta X/\delta A^i\equiv X_{,i}$.
The starting point is a classical action~$S_0(A)$.

The fields are subject to gauge transformations,
\begin{eqnarray}
\delta A^i=R^i_{\alpha}(A)\xi^{\alpha} \qquad{\rm such\ that}\qquad
\delta S_0(A) = 0 \qquad\Leftrightarrow\qquad S_{0,i}(A)R^i_{\alpha}(A)=0,
\end{eqnarray}
where $\xi^{\alpha}$ are arbitrary functions with Grassmann parities
$\varepsilon(\xi^{\alpha})\equiv\varepsilon_{\alpha}$, and
$R^i_{\alpha}(A)$ are the generators of the gauge transformations.
The latter's algebra is inherited from the Lie algebra of the gauge group,
\begin{eqnarray}
\label{GAlgFP}
R^i_{\alpha , j}(A)R^j_{\beta}(A)-(-1)^{\varepsilon_{\alpha}\varepsilon_
{\beta}}R^i_{\beta ,j}(A)R^j_{\alpha}(A)
=- R^i_{\gamma}(A)f^\gamma_{\ \alpha\beta},
\end{eqnarray}
where $f^{\gamma}_{\ \alpha\beta}=
-(-1)^{\varepsilon_{\alpha}\varepsilon_{\beta}}f^{\gamma}_{\ \beta\alpha}$
are the structure constants.
Faddeev-Popov quantization can be applied to the algebra~(\ref{GAlgFP})
if, in addition, the generators $R^i_{\alpha}$ are linearly independent
with respect to~$\{\alpha\}$.

Let us introduce the extended configuration space of fields
as follows:
\begin{eqnarray}
\nonumber
&&\bigl\{\phi^A\bigr\}=\bigl\{A^i,\;B^{\alpha},\;C^{\alpha},\;\bar{C}^{\alpha}\bigr\},
\nonumber\\
&&\varepsilon(A^i)=\varepsilon_i,\;\;
\varepsilon(B^{\alpha})=\varepsilon_{\alpha},\;\;
\varepsilon(C^{\alpha})=
\varepsilon(\bar{C}^{\alpha})=\varepsilon_{\alpha}+1,
\nonumber\\
&&gh(A^i)=gh(B^{\alpha})=0,\quad gh(C^{\alpha})=1, \quad
gh(\bar{C}^{\alpha})=-1,
\nonumber
\end{eqnarray}
where $B^{\alpha}$ are Nakanishi-Lautrup auxiliary fields,
and $C^{\alpha}$ and $\bar{C}^{\alpha}$ are the
Faddeev-Popov ghost and anti-ghost fields, respectively.
We also have introduced the ghost-number grading~$gh$.
Then, the total
action is constructed according to the Faddeev-Popov rule,
\begin{eqnarray}
\label{FPAction} S(\phi)=S_0(A)+\bar{C}^{\alpha}\chi_{\alpha
,i}(A)R^i_{\beta}(A)C^{\beta} +\chi_{\alpha}(A)B^{\alpha}
\end{eqnarray}
where $\chi_{\alpha}(A)$ with
$\varepsilon(\chi_{\alpha}){=}\varepsilon_{\alpha}$
are some gauge functionals which lift the degeneracy of the
classical gauge-invariant action $S_0(A)$.

The generating functional of the Greens functions is written
in the form of a functional integral,
\begin{eqnarray}
\label{FPZ} Z(J)=\int {\cal D}\phi\
\exp\Big\{\frac{\im}{\hbar}\big(S(\phi) +J_A\phi^A\big) \Big\}.
\end{eqnarray}
If, in addition,
\begin{eqnarray}
\label{FPAdR}
(-1)^{\varepsilon_{\beta}}f^{\beta}_{\ \beta\alpha}=0 \und
(-1)^{\varepsilon_i}\frac{\delta_l R^i_{\alpha}}{\delta A^i}=0\ ,
\end{eqnarray}
then one can prove the gauge independence of the vacuum functional $Z(0)$
and of the $S$-matrix.  For pure Yang-Mills theories
the relations (\ref{FPAdR}) are valid due to the
antisymmetry of the structure constants $f^a_{bc}$.

The action (\ref{FPAction}) is invariant under the BRST
transformation \cite{brs,t}
\begin{eqnarray}
\label{FPBRST}
&&\delta_\lambda A^i=R^i_{\alpha}(A)C^{\alpha}\lambda,
\qquad
\delta_\lambda C^{\alpha}=-\sfrac12 (-1)^{\varepsilon_{\beta}}
f^{\alpha}_{\ \beta\gamma}C^{\gamma}C^{\beta}\lambda,
\\
\label{FPBRST2}
&&\delta_\lambda\bar{C}^{\alpha}=B^{\alpha}\lambda,
\qquad \qquad ~
\delta_\lambda B^{\alpha}=0.
\end{eqnarray}
Here, $\lambda$ is a constant Grassmann parameter
($\varepsilon(\lambda)=1$).
Due to the gauge invariance of $S_0$ and the Jacobi identity
for the structure constants, the BRST transformation is nilpotent.
It is quite useful to introduce the Slavnov variation $sX$
of any functional $X$ by writing
\beq
\label{BRSToper}
\delta_\lambda X(\phi) = \bigl(sX(\phi)\bigr)\,\lambda
\qquad\Longrightarrow\qquad
sX(\phi) = \frac{\delta X(\phi)}{\delta\phi^A}R^A(\phi)
\equiv X_{,A}(\phi) R^A(\phi)
\eeq
with the notation
\beq
\label{RA}
R^A(\phi)=\big(R^i_{\alpha}(A)C^{\alpha},\; 0\;, -\sfrac12 (-1)^{\varepsilon_{\beta}}
f^{\alpha}_{\ \beta\gamma}C^{\gamma}C^{\beta},\; B^{\alpha}\big),
\quad \varepsilon(R^A(\phi))=\varepsilon_A+1\ .
\eeq
In particular, from (\ref{FPBRST}) and (\ref{FPBRST2}) we read off $s\,\phi^A$
for all the fields.

With the fermionic gauge-fixing functional
\beq
\label{psi}
\psi(\phi)={\bar C}^{\alpha}\chi_{\alpha}(A)
\eeq
we can present the action (\ref{FPAction}) in the form
\beq
\label{FPAction1}
S(\phi)=S_0(A)+\frac{\delta\psi(\phi)}{\delta A^i}R^i_{\alpha}C^{\alpha}+
\frac{\delta \psi(\phi)}{\delta {\bar C}^{\alpha}}B^{\alpha}
=S_0(A)+s\,\psi(\phi)
\eeq
where its BRST invariance is obvious.
The nilpotency of the Slavnov variation, $s^2=0$, implies that
\beq
\label{nils}
0 = s R^A(\phi) = \frac{\delta R^A(\phi)}{\delta\phi^B}R^B(\phi)
\equiv R^A_{,B}(\phi) R^B(\phi)\ .
\eeq

Let us indicate the choice of the gauge-fixing functional~$\psi$ in the definition
of the generating functional by denoting it as $Z_\psi(J)$.
It is well known that the vacuum functional of Yang-Mills theory does
not depend on the gauge. It has been proven that, under an infinitesimal
change of the gauge-fixing functional, one has
\beq
Z_{\psi}(0)=Z_{\psi+\delta\psi}(0)\ .
\eeq

\vspace{0.5cm}

\section{Field-dependent BRST transformations}

\noindent
In this section we are going to consider a more general class of
BRST transformations by allowing its Grassmann parameter to depend on
the fields of the theory.
Such transformations are admissible in the
functional formulation of quantum field theory.
So let us generalize (\ref{BRSToper}) to
\beq
\label{FDBRST}
\delta_\Lambda X(\phi) = \bigl( sX(\phi)\bigr) \Lambda(\phi)
= X_{,A} R^A \Lambda(\phi), \qquad
\varepsilon(\Lambda(\phi))=1,\quad \Lambda^2(\phi)=0.
\eeq
On the field themselves, a shift by a Slavnov variation amounts to a change of variables,
\beq
\label{FFBRST}
\varphi^A=\varphi^A(\phi)
=\phi^A+\delta_\Lambda\phi^A
=\phi^A+(s\phi^A)\Lambda(\phi)
=\phi^A+R^A(\phi)\Lambda(\phi)\ ,
\eeq
with a Jacobian supermatrix
\beq
\begin{aligned}
M^A_{\ B}(\phi)=\frac{\delta \varphi^A(\phi)}{\delta\phi^B}
&=\delta^A_{\ B}+
\frac{\delta R^A(\phi)}{\delta\phi^B}\Lambda(\phi)(-1)^{\varepsilon_B}+
R^A(\phi)\frac{\delta \Lambda(\phi)}{\delta\phi^B} \\
&\equiv\delta^A_{\ B}+
R^A_{,B}\Lambda(\phi)(-1)^{\varepsilon_B}+R^A\Lambda_{,B}(\phi)\ ,
\end{aligned}
\eeq
where $\varepsilon(M^A_{\ B}(\phi))=\varepsilon_A+\varepsilon_B$.

Now consider a functional integral
\beq
\label{FIW}
{\cal I}=\int {\cal D}\varphi\;\exp\Big\{\frac{\im}{\hbar}W(\varphi)\Big\}
\eeq
with some functional $W(\varphi)$.
Changing the field variables according to (\ref{FFBRST}) then yields
\beq
\label{FIWTr}
\begin{aligned}
{\cal I}&=\int {\cal D}\phi\; \sDet M(\phi)\;
\exp\Big\{\frac{\im}{\hbar}W(\varphi(\phi))\Big\}\\
&=\int {\cal D}\phi\; \exp\Big\{\frac{\im}{\hbar}\big[W(\varphi(\phi))-\im\hbar\,
\sTr\ln M(\phi)\big]\Big\}
\end{aligned}
\eeq
where $\sDet $ and $\sTr $ denote the functional superdeterminant and supertrace,
respectively.
Due to $\Lambda^2=0$ and (\ref{nils}), the computation of $\sTr\ln M$ simplifies
considerably:
\beq
\begin{aligned}
\sTr\ln M(\phi) &=
-\sum_{n=1}^\infty\frac{(-1)^n}{n}\sTr\bigl(
R^A_{,B}\Lambda(-1)^{\varepsilon_B}+R^A\Lambda_{,B}\bigr)^n \\
&=-\sum_{n=1}^\infty\frac{(-1)^n}{n}\sTr\bigl( R^A\Lambda_{,B}\bigr)^n\\
&=+\sum_{n=1}^\infty\frac{(-1)^n}{n}\bigl(\Lambda_{,A} R^A\bigr)^n
= \sum_{n=1}^\infty\frac{(-1)^n}{n}(s\Lambda)^n\\
&=-\ln\bigl(1+s\Lambda(\phi)\bigr).
\end{aligned}
\eeq
Hence, we can give an explicit formula for the Jacobian of an arbitrary
field-dependent BRST transformation,
\beq
\label{sdetformula}
\sDet M(\phi)=\frac{1}{1+s\Lambda(\phi)}.
\eeq
Employing $W(\phi+\delta_\Lambda\phi)=W(\phi)+\delta_\Lambda W(\phi)$,
the functional integral (\ref{FIWTr}) can be expressed as
\beq
\label{Iformula}
{\cal I} =\int {\cal D}\phi\; \exp\Big\{\frac{\im}{\hbar}
\big[W(\phi)+\bigl(sW(\phi)\bigr)\Lambda(\phi)+
\im\hbar\,\ln\bigl(1+s\Lambda(\phi)\bigr)\big]\Big\}
\eeq
being valid for arbitrary functionals $W(\phi)$ and $\Lambda(\phi)$.

In contrast to the standard BRST transformation $\delta_\lambda$,
the field-dependent generalization $\delta_\Lambda$
fails to be nilpotent since, for $sX\neq0$,
\beq
\delta_\Lambda^2 X(\phi)
=\delta_\Lambda\bigl[ \bigl(sX(\phi)\bigr)\Lambda(\phi)\bigr]
=\bigl(sX(\phi)\bigr)\bigl(s\Lambda(\phi)\bigr)\Lambda(\phi)
\eeq
vanishes only if
\beq
0 = s\Lambda(\phi) = \Lambda_{,A}(\phi)R^A(\phi)
\qquad\Longrightarrow\qquad \Lambda(\phi)=\lambda={\rm constant}.
\eeq
In this case, however, $\sDet M(\phi)=1$, and the transformation
should be considered as trivial.

\vspace{0.5cm}

\section{Relating different gauges}

\noindent
Let us apply the results of the previous section to the Yang-Mills vacuum functional
\beq
Z_\psi(0)=\int {\cal D}\varphi\; \exp\Big\{\frac{\im}{\hbar}S(\varphi)\Big\}.
\eeq
Since the action (\ref{FPAction1}) is invariant under the field-dependent
BRST transformation (\ref{FFBRST}), i.e.\ $S(\varphi(\phi))=S(\phi)$,
using the formula~(\ref{Iformula}) yields
\beq
Z_\psi(0)=\int {\cal D}\phi\; \exp\Big\{\frac{\im}{\hbar}
\big[S(\phi)+\im\hbar\,\ln\big(1+s\Lambda(\phi)\big)\big]\Big\}.
\eeq
It may seem strange that (\ref{sdetformula}) may be inserted into the vacuum functional
without any cost, but this becomes clear by writing
\beq
\label{psishift}
\im\hbar\,\ln\big(1+s\Lambda(\phi)\big)= s\,\delta\psi(\phi) \qquad{\rm with}\qquad
\delta\psi(\phi)=
\im\hbar\,\Lambda(\phi)\bigl(s\Lambda(\phi)\bigr)^{-1}\ln\big(1+s\Lambda(\phi)\big).
\eeq
It follows that the insertion of the Jacobian (\ref{sdetformula}) amounts to
adding another BRST-exact piece to the action,
\beq
\label{absorb}
Z_\psi(0)= \int {\cal D}\phi\;\exp\Big\{
\frac{\im}{\hbar}\big[S_0(A)+s\,\psi(\phi)+s\,\delta\psi(\phi)\big]\Big\}
= Z_{\psi+\delta\psi}(0).
\eeq
We see that an arbitrary field-dependent BRST transformation in the vacuum functional
can be transformed into a modification of the gauge-fixing functional,
which keeps the vacuum functional unchanged.
Let us note that this reasoning does not require $\delta\psi$ to be infinitesimal.
It holds not only on a $\delta\psi$-linearized level but {\it exactly\/},
in particular to all orders in a power series expansion in~$\delta\psi$.

It is instructive to turn the above result around:
Any change of gauge, $\psi\to\psi+\delta\psi$, may be effected by a field-dependent
BRST transformation, whose parameter $\Lambda(\phi)$ can be found by inverting
(\ref{psishift}), i.e.~by solving
\beq
\label{invert}
s\Lambda(\phi) = \exp\big\{\sfrac{1}{\im\hbar}s\,\delta\psi\bigr\}-1\ .
\eeq
Up to BRST-exact terms, the solution reads~\footnote{
An analogous formula had been derived differently in~\cite{JM}.}
\beq
\label{solution}
\begin{aligned}
\Lambda(\phi) &\= \delta\psi\,(s\,\delta\psi)^{-1} 
\bigl(\exp\big\{\sfrac{1}{\im\hbar}s\,\delta\psi\bigr\}-1\bigr) \\
&\= \sfrac{1}{\im\hbar}\,\delta\psi\,\sum_{n=0}^\infty \sfrac{1}{(n+1)!}
\bigl(\sfrac{1}{\im\hbar}\,s\,\delta\psi\bigr)^n\ .
\end{aligned}
\eeq

We make this more explicit for changing the parameter $\xi$ in the covariant class of
$R_{\xi}$ gauges, given by the gauge-fixing functional
\beq
\psi(\phi)={\bar C}^a\bigl(\partial^{\mu}A^a_{\mu}+\sfrac{\xi}{2}B^a\bigr) .
\eeq
The field-dependent BRST transformation which connects an $R_\xi$ gauge to
an $R_{\xi+\delta\xi}$ gauge is given by (\ref{invert}) with
\beq
\delta\psi=\sfrac12\delta\xi\,{\bar C}^a\!B^a \qquad\Longrightarrow\qquad
s\,\delta\psi = \sfrac12\delta\xi\,B^2 \qquad{\rm with}\quad B^2=B^a\!B^a\ .
\eeq
{}From (\ref{solution}) we read off that
\beq
\begin{aligned}
\Lambda(\phi) &\= {\bar C}^a\!B^a (B^2)^{-1}
\bigl( \exp\bigl\{\sfrac{\delta\xi}{2\im\hbar}B^2\bigr\}-1 \bigr)\\[4pt]
&\=\sfrac{\delta\xi}{2\im\hbar}\,{\bar C}^a\!B^a \Bigl\{
1+\sfrac{1}{2!}\sfrac{\delta\xi}{2\im\hbar}B^2+
\sfrac{1}{3!}\bigl(\sfrac{\delta\xi}{2\im\hbar}B^2\bigr)^2+
\sfrac{1}{4!}\bigl(\sfrac{\delta\xi}{2\im\hbar}B^2\bigr)^3+\ldots \Bigr\}\ .
\end{aligned}
\eeq
By taking $\delta\xi=\xi$, we may connect in particular the Landau gauge (at $\xi{=}0$)
to any of the $R_\xi$ gauges with the help of a field-dependent BRST transformation.

\vspace{0.5cm}

\section*{Acknowledgments}
\noindent The authors thank I.V. Tyutin for useful discussions of
this paper. This work was supported by the DFG grant LE 838/12-1. The
work of PML is also supported by the LRSS grant 224.2012.2,  by the
Ministry of Education and Science of Russian Federation, project
14.B37.21.0774, by the RFBR grant 12-02-00121 and the RFBR-Ukraine
grant 13-02-90430. He is grateful to the Institute
of Theoretical Physics at Leibniz University for hospitality.



\newpage

\begin {thebibliography}{99}

\bibitem{S}
A.A. Slavnov,
{\it Ward identities in gauge theories},
Theor. Math. Phys.
{\bf 10} (1972) 99.

\bibitem{Taylor}
J.C. Taylor,
{\it Ward identities and charge renormalization of the Yang-Mills field},\\
Nucl. Phys. B {\bf 33} (1971) 436.

\bibitem{BV}
I.A. Batalin and G.A. Vilkovisky,
{\it Gauge algebra and quantization},\\
Phys. Lett. B {\bf 102} (1981) 27;\\
{\it Quantization of gauge theories
with linearly dependent generators},\\
Phys. Rev. D {\bf 28} (1983) 2567.

\bibitem{brs}
C. Becchi, A. Rouet and R. Stora,
{\it Renormalization of the abelian Higgs-Kibble model},\\
Commun. Math. Phys. {\bf 42} (1975) 127.

\bibitem{t}
I.V. Tyutin, {\it Gauge invariance in field theory and statistical
physics in operator formalism}, \\
Lebedev Inst. preprint N 39 (1975), {\tt arXiv:0812.0580}.

\bibitem{FV}
E.S. Fradkin and G.A. Vilkovisky,
{\it Quantization of relativistic systems with constraints},\\
Phys. Lett. B {\bf 55} (1975) 224.

\bibitem{BV5}
I.A. Batalin and G.A. Vilkovisky,\\
{\it Relativistic $S$-matrix of dynamical systems
with boson and fermion constraints},\\
Phys. Lett. B {\bf 69} (1977) 309.

\bibitem{DeWitt}
B.S. DeWitt, {\it Dynamical Theory of Groups and Fields},\\
Gordon and Breach, New York, 1965.

\bibitem{FP}
L.D. Faddeev and V.N. Popov, {\it Feynman diagrams for the Yang-Mills field},\\
Phys. Lett. B {\bf 25} (1967) 29.

\bibitem{JM}
S.D. Joglekar and B.P. Mandal, {\it Finite field-dependent BRS transformations},\\
Phys. Rev. D {\bf 51} (1995) 1919.

\end{thebibliography}
\end{document}